\documentstyle[twoside,fleqn,espcrc2]{article}
\input{psfig}
\newcommand{\AmS}{{\protect\the\textfont2
  A\kern-.1667em\lower.5ex\hbox{M}\kern-.125emS}}

\title{
\vspace*{-55pt}
{\small Proceedings of the 2nd International Conference
on Hyperons, Charm, and Beauty Hadrons} \\
\vspace*{-7pt}
{\small Concordia University, Montreal, Quebec, Canada (27-30 August 1996)} \\
\vspace*{-7pt}
{\small Nuclear Physics B (Proc.~Suppl.) 55A (1997) 221-225}\\
\vspace*{20pt}
Search for Mixing and CP Violation in Charm Decays}

\author{Lucien Cremaldi,\address{Department of Physics and Astronomy, \\ 
        University of Mississippi, \\ 
        University, MS 38677, USA}%
                representing the E791 Collaboration}
       
\begin{document}

\begin{abstract}
Standard model mixing and $CP$ violation are expected to be small in the 
charm sector.
Some argue that only in the presence of new physics do we currently expect to
observe these effects. Fermilab experiment E791 has the largest sample of
reconstructed charm decays recorded to date. We have made an exhaustive search
for $D^0 {\overline D^{\,0}}$ mixing and $CP$ violation in this data.
We present 
the charm mixing limits on the mixing parameter $r_{mix}$ using hadronic and 
semileptonic decays, and also present preliminary limits on the $CP$ asymmetry 
parameter, $A_{CP}$, for Cabibbo suppressed decays. 
\end{abstract}

\maketitle

\section{E791}

E791 is a high statistics charm hadroproduction experiment completed at 
Fermilab in 1992. We recorded 20 billion triggers in 500 GeV $\pi^- N$ 
interactions at the Tagged Photon Spectrometer. Over 200K charm particles 
were reconstructed and are used for further charm analyses\cite{dpf1996}.  
 
The detector\cite{dallas} features 17 planes of silicon microstrip detectors 
placed just behind a segmented Pt and C foil target. Charm decays are detected 
in the air gaps between target foils. Precision tracking of the incoming beam
and a precise location of the secondary decay point allows a measurement of
the proper decay time for charm particles with a resolution of about 
.1$\tau_{D^0}$.   

Particle identification is provided by two multi-cell threshold Cherenkov
counters, giving $\pi /K$ separation in the 6-60 GeV/c momentum range. An
electromagnetic and hadron calorimeter provided good e/$\pi$ separation,
as well as photon identification.  Muons are tagged in a pair of hodoscopes
following steel absorber at the downstream end of the spectrometer.
 
\section{$D^0{\overline D^{\,0}}$ MIXING}

$D^0 \bar D^0$ mixing is expected to be small within the standard
model (SM)\cite{blaylock}. The time-integrated fraction of mixed decays 
relative to Cabibbo favored (CF) decays is experimentally defined as the 
parameter 
$r_{mix} = {{N(D^0 \rightarrow {\overline D^{\,0} \rightarrow 
{\overline f})}}\over {N(D^0 \rightarrow f) }}$,
where $f$ is the final state. SM estimates for the mixing parameter 
$r_{mix}$ are in the range $10^{-7}-10^{-10}$. Any sign of mixing above 
these SM predictions, could be seen as a sign of new physics, although 
long range effects cannot be easily ruled out\cite{wolfy}. 

We look for $D^0{\overline D^{\,0}}$ mixing by searching
for wrong-sign ($ws$) final states ${\overline f}$, in a sample of initially 
tagged $D^0$'s, decaying to right sign ($rs$) final state $f$.
In the limit of small mixing (LSM) this is expressed as
$r_{mix} \simeq {1\over 2}{(({ {\Delta m}\over{\gamma}})^2 +  
({ {\Delta \gamma}\over{2\gamma}})^2  )} $, 
${\Delta m \over \gamma}<<1$ and ${\Delta \gamma \over \gamma }<<1$.
$\Delta m$ and $\Delta \gamma$ are the mass and decay rate
differences between $CP$ eigenstates, and $\gamma$ 
is the average decay rate

Doubly Cabibbo suppressed (DCS) decays can mimic the process.
The total $D^0$ wrong sign decay ($ws$) rate, 
is given (LSM) by, 
$$ \Gamma_{ws}(D^0(t)\rightarrow f)
\simeq [{(\Delta m^2+({\Delta \gamma \over 2})^2})~t^2 $$
\begin{equation}
+ 4~|\rho|^2+2~Re(\rho)\Delta \gamma~t
+ 4 Im(\rho)\Delta m~t]~e^{-\gamma{t}}
\end{equation}
The true mixing ($r_{mix}$), 
DCS interference $Re(\rho)$ and $Im(\rho$), 
and pure DCS ($|\rho|^2$) components can be separated by their 
different time evolutions. Here mixing peaks at 2$\tau_{D^0}$ and is 
sensitive to the tails of the DCS decay time distribution.
In semileptonic final states DCS decays are not allowed and the mixing
rate follows a pure $t^2\times e^{-\gamma{t}}$ \linebreak dependence.

\leftline{\bf 2.1. Hadronic Decay Channels}

We search for mixing in hadronic channels by using the charm decays 
$D^{*+} \rightarrow D^0 \pi^+_b $, then $D^0\rightarrow K^{\mp}\pi^+$, 
$K^{\mp}\pi^+\pi^-\pi^-$\cite{purohit}. The initial charm state is tagged 
by the sign of $\pi_b$, the bachelor pion originating in the primary vertex, 
to form a right
sign ($rs$) and wrong sign ($ws$) sample. A neural network optimization of 
signal-over-root-background, $S/ \sqrt{B}$, in the $rs$ samples is used to 
search for mixing with maximum sensitivity in the $ws$ samples. 
The $Q$-value,
$Q = M(K\pi(\pi\pi)\pi_b)-M(K\pi(\pi\pi))-M(\pi_b)$, and associ\-ated proper 
decay time
spectrum of right sign and wrong sign decays are extracted for each decay.

A maximum likelihood fit (MLF) is performed 
which accounts for the true signal, and background sources: misidentified 
$D\rightarrow \pi\pi,KK$, random pions with real $D^0$'s, 
and random pions with fake $D^0$'s. The $K\pi$ and $K3\pi$ $rs$ and $ws$ data 
is fit simultaneously  for $r_{mix}$ and $\rho$ parameters.  
We assume the residual background under the $rs$ and $ws$ signals is the same
in shape, and fit it to a form determined from combining $D^0$'s and $\pi_b$'s
from different events. We do not assume $CP$ invariance 
when fitting any of the $ws$ rates (mixing, DCS, interference), thus particle 
and antiparticle mixing rates are differentiated. Since it is most 
likely that $CP$ will be violated in the DCS interference term, a special 
fit is performed under this condition. 
We also examine the cases of fits with no interference 
term and no $CP$ violation, and no mixing term, to compare with previous 
experiments. A summary of fit results for the hadronic channels is given in 
Table 1.

\begin{figure}[htb]
\vspace*{2.2truein}
\includegraphics{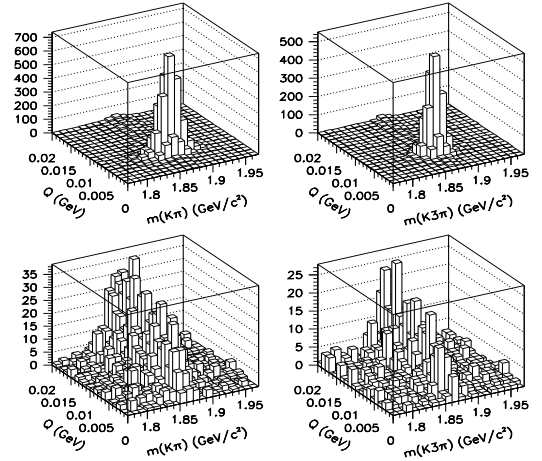}
\caption{Right-sign (upper) and wrong-sign (lower) mixing signals 
for $K\pi$ (left) and $K\pi\pi\pi$ (right) events}
\label{fig:largenenough1}
\end{figure}
\begin{table*}[hbt]
\setlength{\tabcolsep}{1.4pc}
\newlength{\digitwidth} \settowidth{\digitwidth}{\rm 0}
\catcode`?=\active \def?{\kern\digitwidth}
\caption{Results of the MLF to hadronic mixing data described by Eq.1 }
\label{tab:hmix1}
\begin{tabular*}{\textwidth}{@{}l@{\extracolsep{\fill}}rrrr}
\hline
Fit Type      &  Parameter     & E791 Result       &  90\%CL  & Other  \\
\hline

 &  &  &  &  \\ 

Most General,      &     
$r_{mix}(D^0)$ &$(0.70^{+0.58}_{-0.53}\pm 0.18)\%$ & 
$<1.45\%$  &  \\
No CP Assumptions    &     
$r_{mix}({\overline D}^{\,0})$ &$(0.18^{+0.43}_{-0.39}\pm 0.17)\%$ & 
$<0.74\%$ &  \\
 &  &  &  &  \\ 

CP allowed only in       & $r_{mix}$ & $(0.39^{+0.36}_{-0.32}\pm 0.16)\%$ & 
$<0.85\%$   &  \\
Interference Term    &        &        &         &          \\
 &  &  &  &  \\ 

No CP Violation       & $r_{mix}$ & $(0.21^{+0.09}_{-0.09}\pm 0.02)\%$ & 
$<0.33\%$   & $<0.37\%$ \\
No Interference    &        &        &         & E691\cite{anjos88}\\
 &  &  &  &  \\ 

No Mixing      & $r_{DCS}((K\pi)$       &$(0.68^{+0.34}_{-0.33}\pm 0.07)\%$  & 
&$(0.77 \pm 0.35)\%$  \\
               & $r_{DCS}(K\pi\pi\pi)$ & $(0.25^{+0.36}_{-0.34}\pm 0.03)\%$ & 
&  CLEO\cite{cinabro}  \\
 & & & &  \\
\hline
\end{tabular*}
\end{table*}
 
\subsection {Semileptonic Decay Channels}

Charm decays through semileptonic channels offer a unique
opportunity to search for mixing without DCS interference effects 
( $\rho=0$ in eq. 1).
E791 has reported the first measurement of $D^0 \bar D^0$ mixing with 
semileptonic decays, through the decay chain, $D^{*+} \rightarrow D^0 \pi^+_b$, 
then $D^0\rightarrow K^-\ell^+ \bar \nu$\cite{arun96}. Even though there is a 
missing neutrino in the decay, due to the small phase space available, 
an approximate, but narrow, $Q = M(K\ell\pi_b)-M(K\ell)-M(\pi_b)$ distribution 
results, see Figure 2. The momentum of the $D$ can only be determined up to a 
two-fold ambiguity, based on balancing the the momentum of the $D$ meson decay 
products about its line-of-flight, as determined from the vertexing 
information. The shape of the background for the $ws$ signal fits is
determined from combining $D$'s and $\pi_b$'s from different events as in the
hadronic case. A MLF to the $Q$ and decay time distributions is performed.
The results for $r_{mix}$ are summarized in Table 2. 

\begin{figure}[htb]
\vspace*{1.5 truein}
\includegraphics{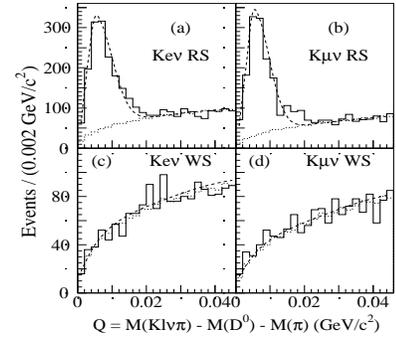}
\caption{Right sign and wrong sign semileptonic decays
used in the mixing analysis. The background and fit projections 
are superimposed.}
\label{fig:largenenough2}
\end{figure}

\begin{table*}[hbt]
\setlength{\tabcolsep}{1.5pc}
\newlength{\ddigitwidth} \settowidth{\ddigitwidth}{\rm 0}
\catcode`?=\active \def?{\kern\ddigitwidth}
\caption{Results of the MLF to semileptonic mixing data described by Eq.1 }
\label{tab:hmix2}
\begin{tabular*}{\textwidth}{@{}l@{\extracolsep{\fill}}rrrr}
\hline
Fit Type      &  Parameter     & E791 Result       &  90\%CL  & Other  \\
\hline

 &  &  &  &  \\ 

Most General,      &     
$r_{mix}(K e \nu)$ &$(0.16^{+0.42}_{-0.37}\pm 0.18)\%$ & 
&  \\
                   &
$r_{mix}(K \mu \nu)$ &$(0.06^{+0.44}_{-0.40}\pm 0.18)\%$ & 
&  \\
Average       &
$r_{mix}(K l \nu)$ &$(0.11^{+0.30}_{-0.27}\pm 0.18)\%$ & 
$<0.50\%$  &  \\

 &  &  &  &  \\ 
\hline
\end{tabular*}
\end{table*}

\section{$CP$ Violation}
\vspace*{-4pt}

$CP$ violating effects are also predicted to be small in the 
standard model\cite{blaylock}. In order for $CP$ violation to occur
there must exist two transition amplitudes to a final state $f$ which interfere
with nonzero relative phase. The situation is reached in $D^0$ decays through
mixing (indirect $CP$ violation, $\Delta C=2$), when interference of 
$CKM$ couplings in 
the relevant amplitudes ($\Psi$) of the box diagrams occurs, $\Psi(D^0 
\rightarrow f)  \neq
\Psi(D^0 \rightharpoonup \bar D^0 \rightarrow f)$. 
This $CP$ violating asymmetry is suppressed by an already low mixing rate,
and is not expected to be seen, $O(10^{-10})$. Direct $CP$ violating effects 
will manifest themselves in decay rate asymmetries of both 
neutral and charged $D$ mesons, $A^{(f)}_{CP}= {\Gamma(D 
\rightarrow f)-\Gamma(\bar D \rightarrow \bar f)} \over{\Gamma(D\rightarrow f) 
+\Gamma(\bar D \rightarrow \bar f)}$. A particle and antiparticle asymmetry 
may be 
produced by final state interactions and penguin terms in Cabibbo suppressed
modes. These $CP$ violating rate asymmetries in singly Cabibbo suppressed (SCS)
$D^0$ and $D^+$ decays may be as high as a few times $10^{-3}$
\cite{goldenb}. Current experiments have not tested $A_{CP}$ beyond the 
$\approx 10^{-1}$ range.

\subsection{Singly Cabibbo Suppressed Searches}
E791 has searched for $CP$ violating effects in the SCS decay channels 
$D^0 \rightarrow K^+K^-, \pi^+\pi^-$ decays, and $D^+ \rightarrow 
K^+K^-\pi^+$ decays. Previous searches in these
channels have yielded null results\cite{anjos91,frabetti,bartelt}.
We have also searched for decay rate asymmetries in 
$D^+ \rightarrow \pi^+\pi^-\pi^+$ decays\cite{copty}, for the first time.  
To remove $D/{\overline D}$ production and reconstruction asymmetries,
we normalize each SCS mode to its CF counterpart ($K^-\pi^+$, $K^-\pi^+\pi^+$)
and reformulate a working definition of the $CP$ asymmetry as
\vspace*{-10pt}
\begin{equation}
\hspace*{26mm} 
A_{CP} = {{\eta(D)-\eta({\overline D})}\over{\eta(D)+\eta({\overline D})}},
\end{equation}
\vspace*{-3pt}
where $\eta(D) =  {{N(D \rightarrow f_{SCS})} \over 
{N(D \rightarrow f_{CF})}}$, is the
ratio of the number of SCS to CF decays.

In the decay $D^+ \rightarrow K^+K^-\pi^+$, we have inspected each subresonance
$\phi \pi^+$, $K^*K$, as well as the nonresonant $KK\pi$ for individual 
$CP$ violating asymmetries. We have investigated 
$D^+ \rightarrow \pi^+\pi^-\pi^+$ decays, inclusively, shown in Figure 3.
In each case the selection criteria for the SCS mode is arrived at by
optimizing $S/\sqrt{S+B}$ where $S$ is the scaled-to-SCS-level CF decay and $B$
is the background for that SCS decay. The  normalizing CF 
mode is subject to the same optimizing cuts as its SCS counterpart to minimize 
systematic uncertainties. Backgrounds from CF mass reflections and particle 
misidentification are also removed. A simultaneous MLF fit for the 
particle/antiparticle yields is performed to arrive at the 
preliminary $CP$ asymmetries listed in Table 3.

\begin{figure}[htb]
\vspace*{1.5truein}
\includegraphics{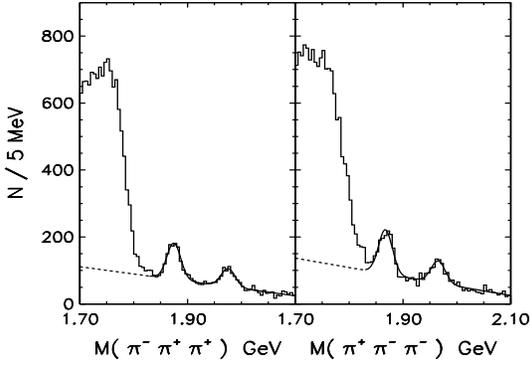}
\caption{$D^+\rightarrow \pi^{\pm} \pi^{\mp}\pi^{\pm}$ events used in the
$CP$ asymmetry calculation. The $K\pi\pi$ reflection below 1.83 GeV was
excluded from the fit.}
\label{fig:largenenough3}
\end{figure}

\begin{table*}[hbt]
\setlength{\tabcolsep}{1.5pc}
\newlength{\dddigitwidth} \settowidth{\dddigitwidth}{\rm 0}
\catcode`?=\active \def?{\kern\dddigitwidth}
\caption{Preliminary CP asymmetry limits set by E791 and other experiments
for CS $D^+$ and $D^0$ decays}
\label{tab:hmix3}
\begin{tabular*}{\textwidth}{@{}l@{\extracolsep{\fill}}rrrr}
\hline
Mode      & $A_{CP}$     &  90\%CL Limit (\%)  & Other  \\
\hline
$K\pi\pi$ &  -  .032             &                       &        \\
$KK\pi$   & $-0.014\pm 0.029$ & $ -6.2 < A_{CP} < +3.4$ 
& $-0.031\pm0.068$ \cite{frabetti} \\
$\phi\pi$ & $-0.028\pm 0.036$ & $-8.7 < A_{CP} < +3.1$        
& $0.066\pm0.086$ \cite{frabetti} \\
${\overline K^{\,*0}}K$ & $-0.010\pm 0.050$ & $ -9.2 < A_{CP} < +7.2$         
& $-0.012\pm0.013$ \cite{frabetti} \\
$\pi\pi\pi$ & $-0.020\pm 0.054$ & $ -8.6 < A_{CP} < +5.2$ &        \\
$KK$     & $-0.017\pm 0.042$ & $ -9.3 < A_{CP} < +6.7$ 
& $0.080\pm0.061$\cite{bartelt}        \\
$\pi\pi$ & $-0.049\pm 0.078$ & $ -17.8 < A_{CP} < +8.0$ &        \\
$K\pi\pi\pi$ & $-0.003\pm 0.021$ & $ -3.6 < A_{CP} < +3.1$ &        \\
\hline
\end{tabular*}
\end{table*}
In the two-body $D^0 \rightarrow \pi^+\pi^-, K^+K^+$ decays, Figure 4, a 
similar approach is taken. 
Here we show the joint $D^0\rightarrow K\pi,KK$, and $\pi\pi$ decay modes
used in the $CP$ asymmetry analysis.
We use SCS $D^0$ decays which are tagged by
the $D^{*+}\rightarrow D^0 \pi_b^+$ and then $D^0\rightarrow K^+K^-,\pi^+\pi^-$
decays. We normalize each SCS yield to the tagged CF counterpart, 
$D^0\rightarrow K^-\pi^+$. Again, we 
remove backgrounds from the CF mass reflections and particle misidentification.
As a consistency check of a null result we also determine $A_{CP}$ for  
the CF mode $D^0\rightarrow K^-\pi^+\pi^+\pi^- $. A MLF fit is performed to
determine the particle/antiparticle yields. We list preliminary results 
for these $CP$ asymmetry measurements, also in Table 3.

\begin{figure}[htb]
\vspace*{2.0truein}
\includegraphics{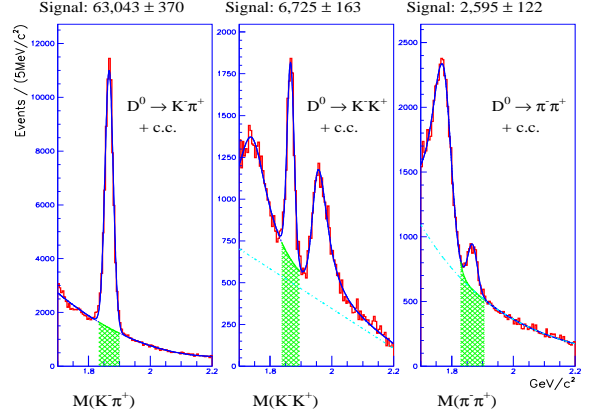}
\caption{$D^0\rightarrow \pi^+ \pi^-, K^+ K^-$ events used in the
$CP$ asymmetry calculation}
\label{fig:largenenough4}
\end{figure}

\section{Conclusion}
E791 has made an exhaustive search for $D^0{\overline D{\,^0}}$ mixing in
Cabibbo favored  hadronic and semileptonic decays. We have also searched for
$CP$ violating asymmetries in singly Cabibbo suppressed modes
including $D^+\rightarrow \pi\pi\pi$ decays for the first time. 
All measurements of mixing and $CP$ 
violation are consistent with zero. With both
hadronic and semileptonic mixing studies we have achieved sensitivities 
on the order of $5\times10^{-3}$ based on general assumptions, possibly testing
standard model extensions. We have improved $CP$ asymmetry measurements 
of SCS decays, in
some cases to the (2-5)\% range, but are not yet approaching a sensitivity 
of $O(10^{-3})$ where $CP$ violating asymmetries are predicted in SCS decays 
in the standard model.

\end{document}